\documentclass[runningheads]{svmult}

\usepackage{epsfig}

\usepackage{makeidx}   % allows index generation
\usepackage{graphicx}  % standard LaTeX graphics tool
\usepackage{subeqnar}  % subnumbers individual equations
\usepackage{multicol}  % used for the two-column index
\usepackage{physmubb}  % centered layout of captions etc.
\makeindex             % used for the subject index
\newcommand{\lsim}{
\mathrel{\hbox{\rlap{\hbox{\lower4pt\hbox{$\sim$}}}\hbox{$<$}}}}
\newcommand{\gsim}{
\mathrel{\hbox{\rlap{\hbox{\lower4pt\hbox{$\sim$}}}\hbox{$>$}}}}

\begin{document}

%%%%%%%%%%% CERN Titlepage %%%%%%%%%%%

\begin{titlepage}
\begin{flushright}
\begin{tabular}{l}
CERN-TH/2002-310\\
hep-ph/0211010\\
November 2002
\end{tabular}
\end{flushright}

\vspace*{1.3truecm}

\begin{center}
\boldmath
{\Large \bf Impact of Bottom-Quark Measurements on\\ 
\vspace*{0.2truecm}
our Knowledge of the Standard Model}
\unboldmath

\vspace*{1.6cm}

\smallskip
\begin{center}
{\sc {\large Robert Fleischer}}\\
\vspace*{2mm}
{\sl Theory Division, CERN, CH-1211 Geneva 23, Switzerland}
%E-mail: {\tt Robert.Fleischer@cern.ch}}
\end{center}

\vspace{2.0truecm}

{\large\bf Abstract\\[10pt]} \parbox[t]{\textwidth}{
In this decade, $B$-decay experiments will allow stringent tests of the 
Standard-Model description of CP violation. After a classification of the 
main strategies and a discussion of the most recent results on benchmark 
decay modes of $B^\pm$ and $B_d$ mesons, we focus on the ``El Dorado'' for 
hadron colliders, the $B_s$-meson system, discussing the differences between 
the $B_d$ and $B_s$ systems, as well as prominent $B_s$ modes. First access 
to these decays is already provided by run II of the Tevatron. In the LHC 
era, it will be possible to fully exploit their physics potential, in 
particular at LHCb and BTeV. 
}

\vspace{1.5cm}
 
{\sl Invited talk at the 14th Topical Conference on Hadron 
Collider Physics,\\
Karlsruhe, Germany, 29 September -- 4 October 2002\\
To appear in the Proceedings}
\end{center}

\end{titlepage}
 
\thispagestyle{empty}
\vbox{}
\newpage
 
\setcounter{page}{1}
 
%%% end CERN title page %%%%%%%%%%%%%

%
%
%
\title*{Impact of Bottom-Quark Measurements
on our Knowledge of the Standard Model}
%
%
%\toctitle{Focusing of a Parallel Beam to Form a Point
%\protect\newline in the Particle Deflection Plane}
% allows explicit linebreak for the table of content
%
%
\titlerunning{Impact of Bottom-Quark Measurements
on our Knowledge of the SM}
% allows abbreviation of title, if the full title is too long
% to fit in the running head
%
\author{Robert Fleischer}
\institute{Theory Division, CERN, CH-1211 Geneva 23, Switzerland}

\authorrunning{Robert Fleischer}
% if there are more than two authors,
% please abbreviate author list for running head
%
%

\maketitle              % typesets the title of the contribution

\section{Introduction}\label{sec:intro}
In 1964, the discovery of CP violation through $K_{\rm L}\to\pi^+\pi^-$ 
decays came as a big surprise 
\cite{CP-discovery}. This particular kind of CP violation, which is 
described by the famous parameter $\varepsilon$, is referred to as 
``indirect'' CP violation, as it is due to the fact that the $K_{\rm L}$ 
mass eigenstate is not an eigenstate of the CP operator with eigenvalue 
$-1$, but receives a tiny admixture of the CP eigenstate with eigenvalue 
$+1$. In 1999, also ``direct'' CP violation, i.e.\ CP-violating effects 
arising directly at the amplitude level, could be established in the 
neutral kaon system by the NA48 (CERN) \cite{NA48} and KTeV (Fermilab) 
collaborations \cite{KTEV}. Unfortunately, the theoretical interpretation 
of the corresponding observable $\mbox{Re}(\varepsilon'/\varepsilon)$ is 
still affected by large hadronic uncertainties and does not provide a 
stringent test of the Standard-Model description of CP violation, unless 
significant theoretical progress concerning the relevant hadronic matrix 
elements can be made \cite{bertolini}--\cite{CiMa}. 

One of the hot topics in this decade is the exploration of decays
of $B$ mesons, allowing powerful tests of the CP-violating sector of 
the Standard Model (SM), and offering valuable insights into hadron dynamics 
\cite{RF-Phys-Rep}. At the moment, the stage is governed 
by the asymmetric $e^+e^-$ $B$ factories operating at the $\Upsilon(4S)$ 
resonance, with their detectors BaBar (SLAC) and Belle (KEK). In 2001, 
these experiments could establish CP violation in the $B$-meson system 
\cite{BaBar-CP-obs,Belle-CP-obs}, which represents the start of a new
era in the exploration of CP violation. Many interesting strategies can 
now be confronted with data \cite{BABAR-BOOK}. In the near future, also 
run II of the Tevatron is expected to contribute significantly to this 
programme, providing -- among other things -- first access to $B_s$-meson 
decays \cite{TEV-BOOK}. In the LHC era, these decay modes can then be fully 
exploited \cite{LHC-BOOK}, in particular at LHCb (CERN) and 
BTeV (Fermilab). 

The focus of this overview is CP violation: in Section~\ref{sec:CP-B},
we give an introduction to the SM description of this 
phenomenon, and classify the main strategies to explore it. In 
Section~\ref{sec:bench}, we shall then have a closer look at important 
benchmark modes of $B^\pm$ and $B_d$ mesons. The ``El Dorado'' for 
$B$-decay studies at hadron colliders, the $B_s$-meson system, is the subject 
of Section~\ref{sec:Bs}, where we shall discuss the differences between 
the $B_d$ and $B_s$ systems, as well as prominent $B_s$ modes. In 
Section~\ref{sec:rare}, we comment briefly on rare $B$ decays, 
before we summarize our conclusions and give an outlook in 
Section~\ref{sec:concl}.

\section{CP Violation in $B$ Decays}\label{sec:CP-B}
\subsection{Weak Decays}
The CP-violating effects we are dealing with in this paper originate from 
the charged-current interactions of the quarks, described by 
\begin{equation}\label{cc-lag2}
{\cal L}_{\mbox{{\scriptsize int}}}^{\mbox{{\scriptsize CC}}}=
-\frac{g_2}{\sqrt{2}}\left(\begin{array}{ccc}\bar
u_{\mbox{{\scriptsize L}}},& \bar c_{\mbox{{\scriptsize L}}},&
\bar t_{\mbox{{\scriptsize L}}}\end{array}\right)\gamma^\mu\,\hat
V_{\mbox{{\scriptsize CKM}}}
\left(
\begin{array}{c}
d_{\mbox{{\scriptsize L}}}\\
s_{\mbox{{\scriptsize L}}}\\
b_{\mbox{{\scriptsize L}}}
\end{array}\right)W_\mu^\dagger \, + \, \mbox{h.c.,}
\end{equation}
where $g_2$ is $SU(2)_{\mbox{{\scriptsize L}}}$ gauge coupling, 
the $W_\mu$ field corresponds to the charged $W$ bosons, and 
$\hat V_{\mbox{{\scriptsize CKM}}}$ denotes the Cabibbo--Kobayashi--Maskawa
(CKM) matrix, connecting the electroweak eigenstates of the down, strange 
and bottom quarks with their mass eigenstates through a unitary 
transformation. 

Since the CKM matrix elements $V_{UD}$ and $V_{UD}^\ast$ enter in 
$D\to U W^-$ and the CP-conjugate process $\overline{D}\to\overline{U}W^+$, 
respectively, where $D\in\{d,s,b\}$ and $U\in\{u,c,t\}$, we observe 
that the phase structure of the CKM matrix is closely related to CP 
violation. It was pointed out by Kobayashi and 
Maskawa in 1973 that actually one complex phase is required -- in addition to 
three generalized Euler angles -- to parametrize the quark-mixing 
matrix in the case of three fermion generations, thereby allowing us to 
accommodate CP violation in the SM \cite{KM}. 

The quark transitions caused by charged-current interactions exhibit
an interesting hierarchy, which is made explicit in the Wolfenstein
parametrization of the CKM matrix \cite{wolf}:
\begin{equation}
\hat V_{\mbox{{\scriptsize CKM}}} =\left(\begin{array}{ccc}
1-\frac{1}{2}\lambda^2 & \lambda & A\lambda^3(\rho-i \eta) \\
-\lambda & 1-\frac{1}{2}\lambda^2 & A\lambda^2\\
A\lambda^3(1-\rho-i \eta) & -A\lambda^2 & 1
\end{array}\right)+{\cal O}(\lambda^4).
\end{equation}
This parametrization corresponds to an expansion in powers of the small 
quantity $\lambda=0.22$, which can be fixed through semileptonic kaon 
decays. The other parameters are of order 1, where $\eta$ leads to an
imaginary part of the CKM matrix. The Wolfenstein parametrization is
very useful for phenomenological applications, as we will see below.

\subsection{Unitarity Triangles}
The central targets for tests of the Kobayashi--Maskawa (KM) mechanism of 
CP violation are the unitarity triangles of the CKM matrix. As we have 
already noted, the CKM matrix is unitary, implying 6 orthogonality 
relations, which can be represented as 6 triangles in the complex 
plane \cite{AKL}, all having the same area \cite{UT-area}. However, 
using the Wolfenstein parametrization, it can be shown that only the 
following two relations describe triangles, where all three sides are 
of the same order of magnitude:
\begin{eqnarray}
V_{ud}\,V_{ub}^\ast+V_{cd}\,V_{cb}^\ast+V_{td}\,V_{tb}^\ast&=&0
\label{UT1}\\
V_{ub}^\ast\, V_{tb}+V_{us}^\ast\, V_{ts}+V_{ud}^\ast\, V_{td}&=&0.\label{UT2}
\end{eqnarray}
At leading order in $\lambda$, these relations agree with each other, and
yield
\begin{equation}\label{UTLO}
(\rho+i\eta)A\lambda^3+(-A\lambda^3)+(1-\rho-i\eta)A\lambda^3=0.
\end{equation}
Consequently, they describe the same triangle, which is usually referred to 
as {\it the} unitarity triangle of the CKM matrix \cite{UT-area,CK}. 
It is convenient to divide (\ref{UTLO}) by the overall normalization 
$A\lambda^3$. Then we obtain a triangle in the complex plane with a 
basis normalized to 1, and an apex given by $(\rho,\eta)$.

\begin{figure}[t]
\begin{tabular}{lr}
   \epsfysize=3.4cm
   \epsffile{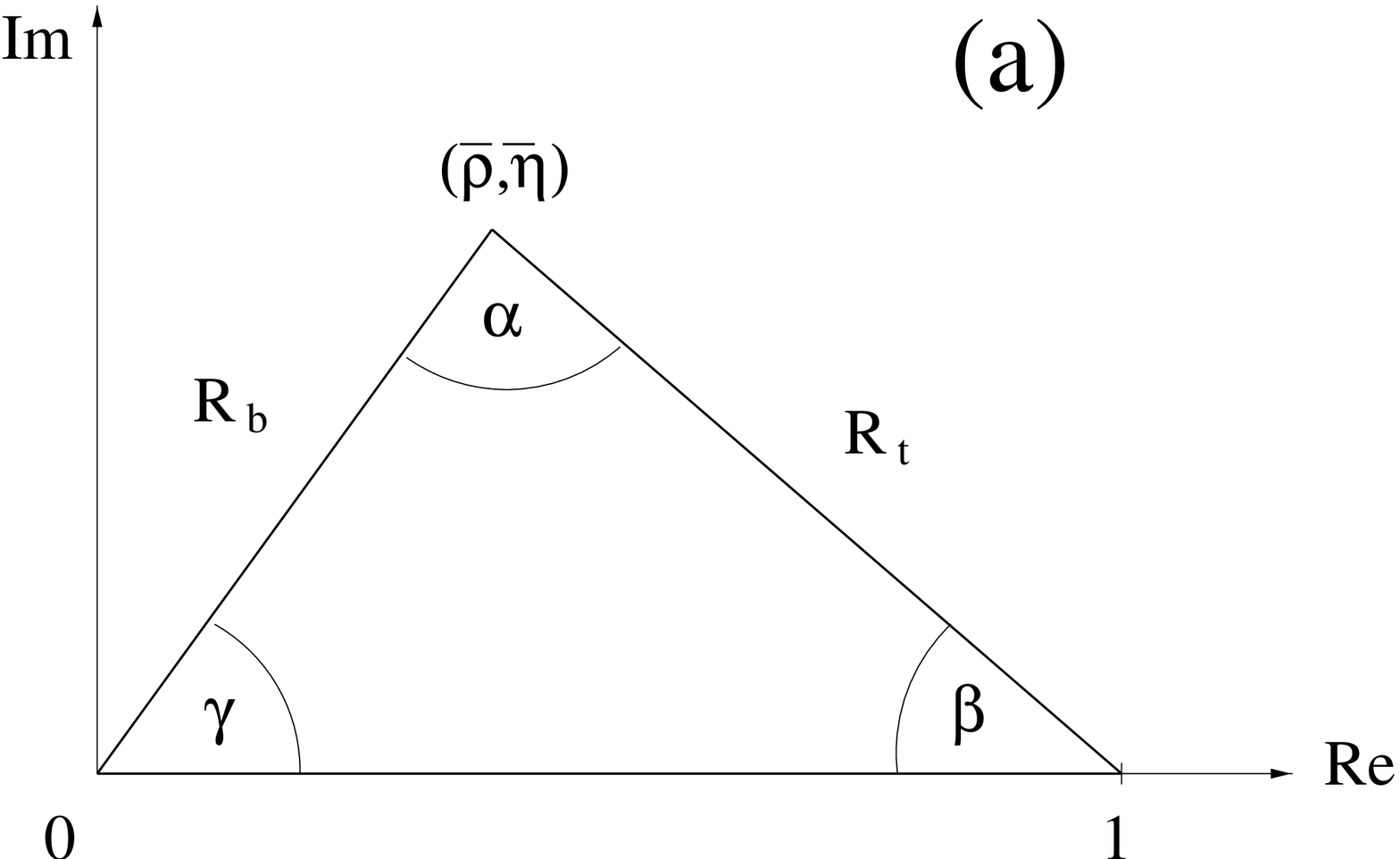}
&
   \epsfysize=3.4cm
   \epsffile{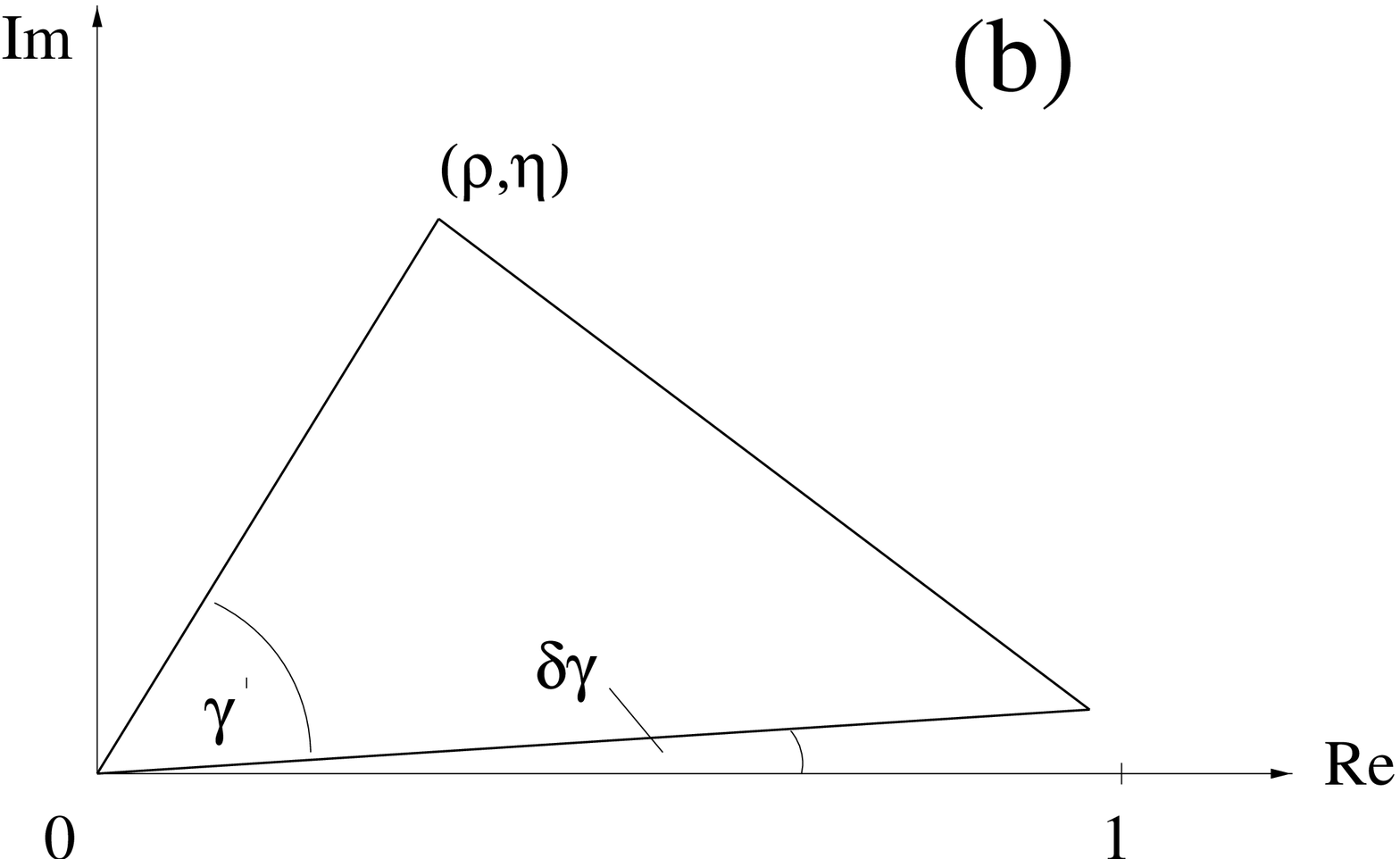}
\end{tabular}
\caption[]{The two non-squashed unitarity triangles of the CKM matrix: 
(a) and (b) correspond to the orthogonality relations (\ref{UT1}) and 
(\ref{UT2}), respectively.}
\label{fig:UT}
\end{figure}

In the future, the experimental accuracy will reach such an impressive 
level that we will have to distinguish between the unitarity triangles 
described by (\ref{UT1}) and (\ref{UT2}), which differ through 
${\cal O}(\lambda^2)$ corrections. They are illustrated in 
Fig.\ \ref{fig:UT}, where $\overline{\rho}$ and 
$\overline{\eta}$ are related to $\rho$ and $\eta$ through \cite{blo}
\begin{equation}
\overline{\rho}\equiv\left(1-\lambda^2/2\right)\rho,\quad
\overline{\eta}\equiv\left(1-\lambda^2/2\right)\eta,
\end{equation}
and 
\begin{equation}
\delta\gamma\equiv\gamma-\gamma'=\lambda^2\eta.
\end{equation}
Whenever we refer to a unitarity triangle, we mean the one shown
in Fig.\ \ref{fig:UT} (a). To determine the allowed region in the 
$\overline{\rho}$--$\overline{\eta}$ plane, the ``standard analysis'' 
uses the following ingredients (for explicit expressions, see
\cite{BF-rev}):
\begin{itemize}
\item Exclusive and inclusive semileptonic $B$ decays caused by
$b\to c\ell\overline{\nu}_{\ell}, u\ell\overline{\nu}_{\ell}$ quark-level
transitions, fixing a circle of radius $R_b$ around $(0,0)$ 
\cite{ligeti}.
\item $B^0_q$--$\overline{B^0_q}$ mixing ($q\in\{d,s\}$), fixing a circle 
of radius $R_t$ around $(1,0)$.
\item Indirect CP violation in the neutral kaon system, $\varepsilon$, fixing
a hyperbola.
\end{itemize}
Many different strategies to deal with the corresponding theoretical and 
experimental uncertainties can be found in the literature. The most 
important ones are the simple scanning approach \cite{buras-KAON}, 
the Gaussian approach \cite{al}, the BaBar 95\% scanning method
\cite{Babar-95-scan}, the Bayesian approach \cite{bayes}, and the
non-Bayesian statistical approach developed in \cite{hoeck}. A detailed 
discussion of these approaches is beyond the scope of this presentation. 
Let us here just give typical ranges for $\alpha$, $\beta$ and $\gamma$ 
that are implied by these strategies:
\begin{equation}\label{UT-Fit-ranges}
70^\circ\lsim\alpha\lsim 130^\circ, \quad
20^\circ\lsim\beta\lsim 30^\circ, \quad
50^\circ\lsim\gamma\lsim 70^\circ.
\end{equation}

Direct determinations of these angles are provided by CP-violating effects
in $B$ decays. The goal is now to overconstrain the unitarity triangle as 
much as possible through independent measurements of its sides and angles, 
with the hope to encounter discrepancies, which may shed light on new 
physics.

\subsection{Main Strategies}
The main r\^ole in the exploration of CP violation through $B$ decays
is played by non-leptonic transitions, as CP-violating effects are due
to interference effects, which arise in this decay class. In particular, 
interference between different decay topologies, i.e.\ tree and penguin 
contributions, may lead to direct CP violation. Unfortunately, the
corresponding CP asymmetries are affected by hadronic matrix elements
of local four-quark operators, which are hard to estimate and preclude
a clean determination of weak phases. In order to solve this problem, 
we may employ one of the following approaches:
\begin{itemize}
\item The most obvious -- but also most challenging -- strategy 
we may follow is to try to calculate the relevant hadronic matrix 
elements $\langle \overline{f}|Q_k(\mu)|\overline{B}\rangle$. Interesting 
progress has recently been made in this direction through the development 
of the QCD factorization \cite{BBNS1}--\cite{fact-recent}, the 
perturbative hard-scattering (PQCD) \cite{PQCD}, and QCD light-cone 
sum-rule approaches \cite{sum-rules}. 
\item Another avenue we may follow is to search for fortunate cases where
relations between decay amplitudes allow us to eliminate the hadronic
matrix elements. Here we distinguish between exact relations, involving
pure tree decays of the kind $B\to KD$ \cite{gw}--\cite{ADS} or 
$B_c\to D_sD$ \cite{FW}, and relations, which follow from the flavour 
symmetries of strong interactions, involving $B_{(s)}\to \pi\pi, \pi K, KK$ 
decays \cite{GRL}--\cite{GR-BpiK-recent}.  
\item The third avenue we may follow to deal with the problems arising 
from hadronic matrix elements is to employ decays of neutral $B_d$ or
$B_s$ mesons. Here we encounter a new kind of CP violation, which is
due to interference effects between $B^0_q$--$\overline{B^0_q}$ 
($q\in\{d,s\}$) mixing and decay processes; it is referred to as 
``mixing-induced'' CP violation. In the rate asymmetry
\begin{eqnarray}
\lefteqn{\left.\frac{\Gamma(B^0_q(t)\to f)-
\Gamma(\overline{B^0_q}(t)\to \overline{f})}{\Gamma(B^0_q(t)\to f)+
\Gamma(\overline{B^0_q}(t)\to \overline{f})}\right|_{\Delta\Gamma_q=0}}
\nonumber\\
&&={\cal A}_{\rm CP}^{\rm dir}(B_q\to f)\cos(\Delta M_q t)+
{\cal A}_{\rm CP}^{\rm mix}(B_q\to f)\sin(\Delta M_q t),
\end{eqnarray}
where $\Delta M_q$ and $\Delta\Gamma_q$ are the $B_q$ mass and decay widths
differences, respectively, and $(CP)|f\rangle=\pm|f\rangle$, it is described 
by the coefficient of the $\sin(\Delta M_q t)$ term, whereas the one of 
$\cos(\Delta M_q t)$ measures direct CP violation. If the decay
$B_q\to f$ is dominated by a single CKM amplitude, the corresponding 
hadronic matrix element cancels in ${\cal A}_{\rm CP}^{\rm mix}(B_q\to f)$.
This observable is then simply given by $\pm\sin(\phi_q-\phi_f)$, where 
$\phi_f$ and $\phi_q$ are the weak $B_q\to f$ decay and 
$B^0_q$--$\overline{B^0_q}$ mixing phases, respectively \cite{RF-Phys-Rep}. 
\end{itemize}

\section{Benchmark Decay Modes of $B^\pm$ and $B_d$ Mesons}\label{sec:bench}
\subsection{$B\to\pi K$}
These decays, which originate from $\overline{b}\to 
\overline{d}d\overline{s},\overline{u}u\overline{s}$ quark-level transitions, 
may receive contributions from penguin and tree 
topologies, where the latter bring the CKM angle $\gamma$ into the game. 
Interestingly, because of $|V_{us}V_{ub}^\ast/(V_{ts}V_{tb}^\ast)|
\approx0.02$, $B\to\pi K$ modes are dominated by QCD penguins, despite 
their loop suppression. As far as electroweak (EW) penguins are concerned, 
they are expected to be negligible in $B^0_d\to\pi^-K^+$, $B^+\to\pi^+K^0$, 
as they contribute here only in colour-suppressed form. On the other hand, 
they are sizeable in $B^+\to\pi^0K^+$ and $B^0_d\to\pi^0K^0$, i.e.\ of 
the same order of magnitude as the trees, since they contribute here also 
in colour-allowed form. 

Through interference effects between tree and penguin contributions, 
we obtain sensitivity on $\gamma$. Relations between the $B\to\pi K$ 
amplitudes that are implied by the $SU(2)$ isospin flavour symmetry of 
strong interactions suggest the following combinations to determine
this angle: the ``mixed'' $B^\pm\to\pi^\pm K$, $B_d\to\pi^\mp K^\pm$ 
system \cite{PAPIII}--\cite{defan}, the ``charged'' $B^\pm\to\pi^\pm K$, 
$B^\pm\to\pi^0K^\pm$ system \cite{NR}--\cite{BF-neutral1}, and 
the ``neutral'' $B_d\to\pi^0 K$, $B_d\to\pi^\mp K^\pm$ system 
\cite{BF-neutral1,BF-neutral2}. 

All three $B\to\pi K$ systems can be described by the same set of 
formulae by just making straightforward replacements of variables 
\cite{BF-neutral1}. Let us here, for simplicity, focus on the charged 
$B\to\pi K$ system. In order to determine $\gamma$ and strong 
parameters, we have to introduce appropriate CP-conserving and
CP-violating observables, which are given as follows:
\begin{equation}\label{charged-obs}
\left\{\begin{array}{c}R_{\rm c}\\A_0^{\rm c}\end{array}\right\}
\equiv2\left[\frac{\mbox{BR}(B^+\to\pi^0K^+)\pm
\mbox{BR}(B^-\to\pi^0K^-)}{\mbox{BR}(B^+\to\pi^+K^0)+
\mbox{BR}(B^-\to\pi^-\overline{K^0})}\right].
\end{equation}
To parametrize these observables, we make use of the isospin relation 
mentioned above, and assume that certain rescattering
effects are small, which is in accordance with the QCD factorization
picture \cite{BBNS1}--\cite{BBNS3}. Anomalously large rescattering processes
would be indicated by data on $B\to KK$ modes, which are already highly
constrained by the $B$ factories, and could be taken into account through
more elaborate strategies \cite{defan,neubert-BpiK,BF-neutral1}. The 
expressions for $R_{\rm c}$ and $A_0^{\rm c}$ thus obtained involve 
then -- in addition to $\gamma$ -- the parameters $r_{\rm c}$, $q$ and 
$\delta_{\rm c}$, which have the following physical interpretation:
$r_{\rm c}$ measures, simply speaking, the ratio of tree to penguin 
topologies. It can be fixed through $SU(3)$ arguments and data on 
$B^\pm\to\pi^\pm\pi^0$ \cite{GRL}, yielding $r_{\rm c}\sim0.2$. On the 
other hand, $q$ describes the ratio of EW penguin to tree contributions, 
and can be determined through $SU(3)$ arguments, yielding $q\sim 0.7$ 
\cite{NR}. Finally, $\delta_{\rm c}$ is the CP-conserving strong phase 
between trees and penguins.

\begin{figure}[t]
$$\hspace*{-1.cm}
\epsfysize=0.19\textheight
\epsfxsize=0.29\textheight
\epsffile{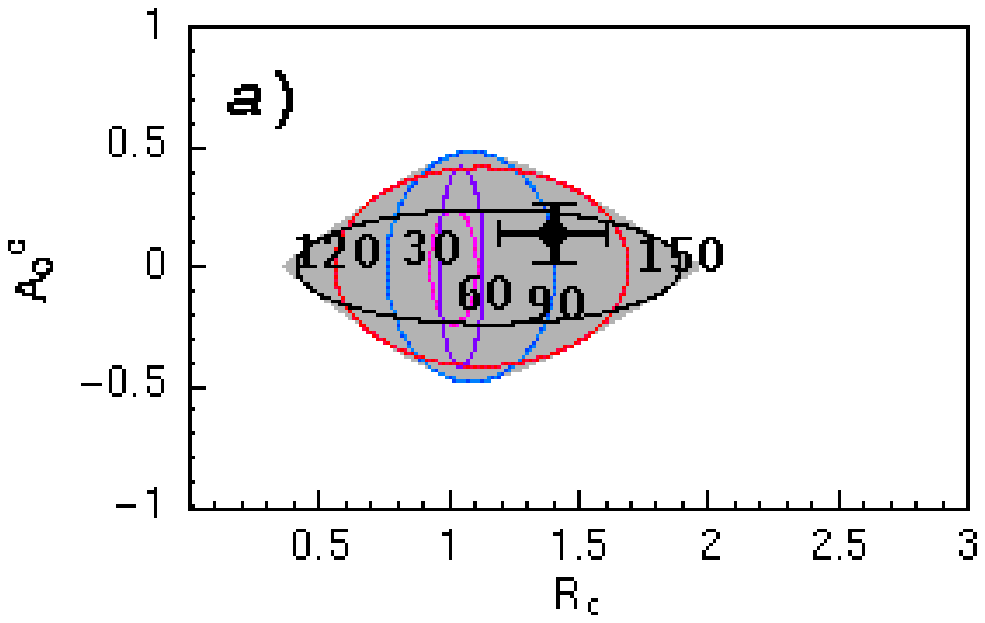} \hspace*{0.3cm}
\epsfysize=0.19\textheight
\epsfxsize=0.29\textheight
\epsffile{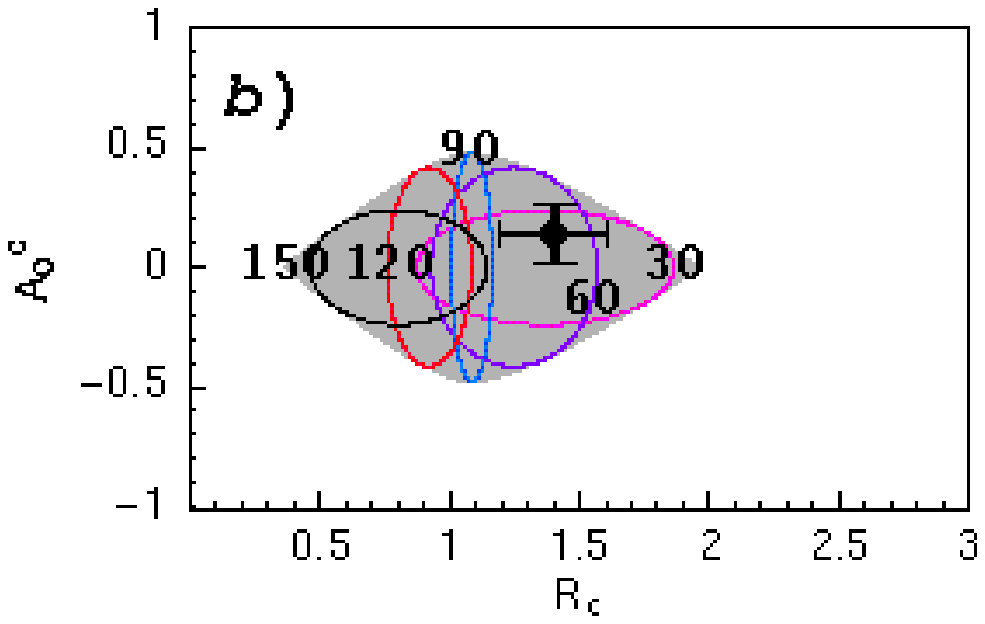}
$$
\caption[]{The allowed regions in the $R_{\rm c}$--$A_0^{\rm c}$ plane
for $q=0.68$ and $r_{\rm c}=0.24$. In (a) and (b), we show also the
contours for fixed values of $\gamma$ and $|\delta_{\rm c}|$, 
respectively.}\label{fig:BpiK-charged}
\end{figure}

Consequently, the two observables $R_{\rm c}$ and $A_0^{\rm c}$ depend on 
the two ``unknowns'' $\delta_{\rm c}$ and $\gamma$. If we vary them within 
their allowed ranges, i.e.\ $-180^\circ\leq \delta_{\rm c}\leq+180^\circ$ 
and $0^\circ\leq \gamma \leq180^\circ$, we obtain an allowed region in 
the $R_{\rm c}$--$A_0^{\rm c}$ plane \cite{FlMa1,FlMa2}. Should the
measured values of $R_{\rm c}$ and $A_0^{\rm c}$ lie outside
this region, we would have an immediate signal for new physics. On the
other hand, should the measurements fall into the allowed range,
$\gamma$ and $\delta_{\rm c}$ could be extracted. In this case, $\gamma$
could be compared with the results of alternative strategies and with
the values implied by the ``standard analysis'' of the unitarity triangle, 
whereas $\delta_{\rm c}$ provides valuable insights into hadron dynamics, 
thereby allowing tests of theoretical predictions. 

In Fig.~\ref{fig:BpiK-charged}, we show the allowed regions in the 
$R_{\rm c}$--$A_0^{\rm c}$ plane for various parameter sets 
\cite{FlMa2}. The crosses represent the averages of the present
$B$-factory data. The contours in Figs.~\ref{fig:BpiK-charged} (a) and (b) 
allow us to read off straightforwardly the preferred values for $\gamma$ 
and $\delta_{\rm c}$, respectively, from the measured observables. 
Interestingly, the present data seem to favour 
$\gamma\gsim90^\circ$ (see also \cite{neubert-02}), which would be in 
conflict with (\ref{UT-Fit-ranges}). Moreover, they point towards 
$|\delta_{\rm c}|\lsim90^\circ$; factorization predicts 
$\delta_{\rm c}$ to be close to $0^\circ$ \cite{BBNS3}. If future, 
more accurate data should really yield a value for $\gamma$ in the second
quadrant, the discrepancy with (\ref{UT-Fit-ranges}) may be due to new-physics
contributions to $B^0_q$--$\overline{B^0_q}$ mixing ($q\in\{d,s\}$) or the
$B\to\pi K$ decay amplitudes. The allowed regions and contours in observable
space of the neutral $B\to\pi K$ system look very similar to those shown in 
Fig.~\ref{fig:BpiK-charged} \cite{FlMa2}; for a recent update, see
\cite{RF-dubna}. Unfortunately, the experimental situation in the neutral 
$B\to\pi K$ system is still rather unsatisfactory. As far as the mixed
$B\to\pi K$ system is concerned, the present data fall well into the
SM region in observable space, but do not yet allow us to 
draw further definite conclusions \cite{FlMa2}. At present, the situation 
in the charged and neutral $B\to\pi K$ systems appears to be more exciting.
Examples of the many other recent $B\to\pi K$ analyses can be found in 
\cite{BBNS3,matias,ital-corr,GR-BpiK-recent,PQCD-appl}.

\subsection{$B\to J/\psi K$}
The decay $B_d^0\to J/\psi\,K_{\rm S}$ is a transition into a CP-odd 
eigenstate, and originates from $\overline{b}\to\overline{c}c\overline{s}$ 
quark-level decays. Consequently, it receives contributions both from tree 
and from penguin topologies. Within the SM, we may write 
\begin{equation}\label{BdpsiK-ampl2}
A(B_d^0\to J/\psi K_{\rm S})\propto\left[1+\lambda^2 a e^{i\theta}
e^{i\gamma}\right],
\end{equation}
where the CP-conserving hadronic parameter $a e^{i\theta}$ measures, 
sloppily speaking, the ratio of the penguin to tree contributions in 
$B_d^0\to J/\psi\,K_{\rm S}$ \cite{RF-BdsPsiK}. Since this parameter 
enters in a doubly Cabibbo-suppressed way, we obtain to a very good 
approximation \cite{bisa} (for a detailed discussion, see 
\cite{RF-Phys-Rep}):
\begin{equation}\label{BpsiK-CP}
{\cal A}_{\rm CP}^{\rm dir}(B_d\to J/\psi K_{\rm S})=0, \quad
{\cal A}_{\rm CP}^{\rm mix}(B_d\to J/\psi K_{\rm S})=-\sin\phi_d,
\end{equation}
where $\phi_d$ denotes the CP-violating weak $B^0_d$--$\overline{B^0_d}$
mixing phase, which is given by $2\beta$ in the SM. After important first 
steps by the OPAL, CDF and ALEPH collaborations, the 
$B_d\to J/\psi K_{\rm S}$ mode (and similar decays) led eventually, 
in 2001, to the observation of CP violation in the $B$ system 
\cite{BaBar-CP-obs,Belle-CP-obs}. The present status of $\sin2\beta$ is 
given as follows:
\begin{equation}
\sin2\beta=\left\{\begin{array}{ll}
0.741\pm 0.067  \pm0.033  &
\mbox{(BaBar \cite{Babar-s2b-02})}\\
0.719\pm 0.074  \pm0.035  &
\mbox{(Belle \cite{Belle-s2b-02}),}
\end{array}\right.
\end{equation}
yielding the world average \cite{nir} 
\begin{equation}\label{s2b-average}
\sin2\beta=0.734\pm0.054, 
\end{equation}
which agrees well with the results of the ``standard analysis'' of the
unitarity triangle (\ref{UT-Fit-ranges}), implying 
$0.6\lsim\sin2\beta\lsim0.9$. 

In the LHC era, the experimental accuracy of the measurement of $\sin2\beta$ 
may be increased by one order of magnitude \cite{LHC-BOOK}. In view of
such a tremendous accuracy, it will then be important to obtain deeper
insights into the theoretical uncertainties affecting (\ref{BpsiK-CP}),
which are due to the penguin contributions described by $a e^{i\theta}$. 
A possibility to control them is provided by the $B_s\to J/\psi K_{\rm S}$ 
channel \cite{RF-BdsPsiK}. Moreover, also direct CP violation in 
$B\to J/\psi K$ modes allows us to probe such penguin effects 
\cite{RF-rev,FM-BpsiK}. So far, there are no experimental indications for 
non-vanishing CP asymmetries of this kind.

Although the agreement between (\ref{s2b-average}) and the results of the
CKM fits is striking, it should not be forgotten that new physics may 
nevertheless hide in ${\cal A}_{\rm CP}^{\rm mix}(B_d\to J/\psi K_{\rm S})$. 
The point is that the key quantity is actually $\phi_d$, which is fixed 
through $\sin\phi_d=0.734\pm0.054$ up to a twofold ambiguity,
\begin{equation}\label{phid-det}
\phi_d=\left(47^{+5}_{-4}\right)^\circ \, \lor \,
\left(133^{+4}_{-5}\right)^\circ.
\end{equation}
Here the former solution would be in perfect agreement with the range
implied by the CKM fits, $40^\circ\lsim\phi_d\lsim60^\circ$, whereas
the latter would correspond to new physics. The two solutions can
be distinguished through a measurement of the sign of $\cos\phi_d$: 
in the case of $\cos\phi_d=+0.7>0$, we would conclude
$\phi_d=47^\circ$, whereas $\cos\phi_d=-0.7<0$ would point towards
$\phi_d=133^\circ$, i.e.\ new physics. 
There are several strategies on the market to resolve the twofold
ambiguity in the extraction of $\phi_d$ \cite{ambig}. Unfortunately,
they are rather challenging from a practical point of view. In the 
$B\to J/\psi K$ system, $\cos\phi_d$ can be extracted from the 
time-dependent angular distribution of the decay products of 
$B_d\to J/\psi[\to\ell^+\ell^-] K^\ast[\to\pi^0K_{\rm S}]$, if the sign 
of a hadronic parameter $\cos\delta$ involving a strong phase $\delta$ 
is fixed through factorization \cite{DDF2,DFN}. This analysis is already 
in progress at the $B$ factories \cite{itoh}. For hadron colliders, the
$B_d\to J/\psi\rho^0$, $B_s\to J/\psi\phi$ system is interesting to probe
$\cos\phi_d$ \cite{RF-ang}.

The preferred mechanism for new physics to manifest itself in CP-violating
effects in $B_d\to J/\psi K_{\rm S}$ is through $B^0_d$--$\overline{B^0_d}$
mixing, which arises in the SM from the famous box diagrams. 
However, new physics may also enter at the $B\to J/\psi K$ amplitude level. 
Employing estimates borrowed from effective field theory suggests that the 
effects are at most ${\cal O}(10\%)$ for a generic new-physics scale 
$\Lambda_{\rm NP}$ in the TeV regime. In order to obtain the whole picture, 
a set of appropriate observables can be introduced, using 
$B_d\to J/\psi K_{\rm S}$ and its charged counterpart 
$B^\pm\to J/\psi K^\pm$ \cite{FM-BpsiK}. So far, these observables do not 
indicate any deviation from the SM.

\subsection{$B\to\phi K$}
Another important testing ground for the KM mechanism of CP violation
is provided by $B\to \phi K$ decays, originating from
$\overline{b}\to\overline{s}s\overline{s}$ quark-level transitions.
These modes are governed by QCD penguins \cite{BphiK-old}, but also 
EW penguins are sizeable \cite{RF-EWP,DH-PhiK}. Consequently, $B\to\phi K$ 
modes represent a sensitive probe for new physics. In the SM, we 
have the relations \cite{RF-rev,growo,loso,FM-BphiK}
\begin{eqnarray}
{\cal A}_{\rm CP}^{\rm dir}(B_d\to \phi K_{\rm S})&=&0+{\cal O}(\lambda^2)\\
{\cal A}_{\rm CP}^{\rm mix}(B_d\to \phi K_{\rm S})&=&
{\cal A}_{\rm CP}^{\rm mix}(B_d\to J/\psi K_{\rm S})+{\cal O}(\lambda^2).
\end{eqnarray}
As in the case of the $B\to J/\psi K$ system, a combined analysis of
$B_d\to \phi K_{\rm S}$, $B^\pm \to \phi K^\pm$ modes should be 
performed in order to obtain the whole picture \cite{FM-BphiK}. There is 
also the possibility of an unfortunate case, where new physics cannot be
distinguished from the SM, as discussed in 
\cite{RF-Phys-Rep,FM-BphiK}. 

In the autumn of 2002, the experimental status can be summarized as 
\begin{equation}
{\cal A}_{\rm CP}^{\rm dir}(B_d\to \phi K_{\rm S})=
\left\{\begin{array}{ll}
\mbox{n.a.} &\mbox{(BaBar \cite{BphiK-BaBar})}\\
0.56\pm0.41\pm0.12
&\mbox{(Belle \cite{BphiK-Belle})}
\end{array}\right.
\end{equation}
\begin{equation}
{\cal A}_{\rm CP}^{\rm mix}(B_d\to \phi K_{\rm S})=
\left\{\begin{array}{ll}
0.19^{+0.50}_{-0.52}\pm 0.09 &\mbox{(BaBar \cite{BphiK-BaBar})}\\
0.73\pm0.64\pm0.18 &\mbox{(Belle \cite{BphiK-Belle}).}
\end{array}\right.
\end{equation}
Unfortunately, the experimental uncertainties are still very large.
Because of ${\cal A}_{\rm CP}^{\rm mix}(B_d\to J/\psi K_{\rm S})=
-0.734 \pm 0.054$ (see (\ref{BpsiK-CP}) and (\ref{s2b-average})), 
there were already speculations about new-physics effects in 
$B_d\to\phi K_{\rm S}$ \cite{BPhiK-NP}. In this context, it is interesting
to note that there are more data available from Belle:
\begin{eqnarray}
{\cal A}_{\rm CP}^{\rm dir}(B_d\to \eta' K_{\rm S})&=&-0.26\pm0.22\pm0.03\\
{\cal A}_{\rm CP}^{\rm mix}(B_d\to \eta' K_{\rm S})
&=&-0.76\pm0.36^{+0.06}_{-0.05}
\end{eqnarray}
\begin{eqnarray}
{\cal A}_{\rm CP}^{\rm dir}(B_d\to K^+K^-K_{\rm S})&=&
0.42\pm0.36\pm0.09^{+0.22}_{-0.03}\\
{\cal A}_{\rm CP}^{\rm mix}(B_d\to K^+K^-K_{\rm S})&=&
-0.52\pm0.46\pm0.11^{+0.03}_{-0.27}.
\end{eqnarray}
The corresponding modes are governed by the same quark-level transitions 
as $B_d\to\phi K_{\rm S}$. Consequently, it is probably too early to get
too excited by the possibility of signals of new physics in 
$B_d\to\phi K_{\rm S}$ \cite{nir}. However, the experimental situation 
should improve significantly in the future.

\subsection{$B\to\pi\pi$}
Another benchmark mode for the $B$ factories is $B_d^0\to\pi^+\pi^-$,
which is a decay into a CP eigenstate with eigenvalue $+1$, and 
originates from $\overline{b}\to\overline{u} u \overline{d}$ 
quark-level transitions. In the SM, we may write 
\begin{equation}\label{Bpipi-ampl}
A(B_d^0\to\pi^+\pi^-)\propto\left[e^{i\gamma}-de^{i\theta}\right],
\end{equation}
where the CP-conserving strong parameter $d e^{i\theta}$
measures, sloppily speaking, the ratio of penguin to tree
contributions in $B_d\to\pi^+\pi^-$ \cite{RF-BsKK}. In contrast to the 
$B_d^0\to J/\psi K_{\rm S}$ amplitude (\ref{BdpsiK-ampl2}), this parameter 
does {\it not} enter in (\ref{Bpipi-ampl}) in a doubly Cabibbo-suppressed 
way, thereby leading to the well-known ``penguin problem'' in 
$B_d\to\pi^+\pi^-$. If we had negligible penguin contributions, i.e.\ 
$d=0$, the corresponding CP-violating observables were given as follows:
\begin{equation}
{\cal A}_{\rm CP}^{\rm dir}(B_d\to\pi^+\pi^-)=0, \quad
{\cal A}_{\rm CP}^{\rm mix}(B_d\to\pi^+\pi^-)=\sin(2\beta+2\gamma)=
-\sin 2\alpha,
\end{equation}
where we have also used the unitarity relation $2\beta+2\gamma=2\pi-2\alpha$. 
We observe that actually the phases $\phi_d=2\beta$ and $\gamma$ 
enter directly in the $B_d\to\pi^+\pi^-$ observables, and not $\alpha$. 
Consequently, since $\phi_d$ can be fixed straightforwardly through 
$B_d\to J/\psi K_{\rm S}$, we may use $B_d\to\pi^+\pi^-$ to probe 
$\gamma$ \cite{FlMa2}. 

Measurements of the $B_d\to\pi^+\pi^-$ CP asymmetries are already 
available:
\begin{equation}\label{Adir-exp}
{\cal A}_{\rm CP}^{\rm dir}(B_d\to\pi^+\pi^-)=\left\{
\begin{array}{ll}
-0.30\pm0.25\pm0.04 & \mbox{(BaBar \cite{BaBar-Bpipi})}\\
-0.94^{+0.31}_{-0.25}\pm0.09 & \mbox{(Belle \cite{Belle-Bpipi})}
\end{array}
\right.
\end{equation}
\begin{equation}\label{Amix-exp}
{\cal A}_{\rm CP}^{\rm mix}(B_d\to\pi^+\pi^-)=\left\{
\begin{array}{ll}
-0.02\pm0.34\pm0.05& \mbox{(BaBar \cite{BaBar-Bpipi})}\\
1.21^{+0.27+0.13}_{-0.38-0.16} & \mbox{(Belle \cite{Belle-Bpipi}).}
\end{array}
\right.
\end{equation}
Unfortunately, the BaBar and Belle results are not fully consistent with
each other; the experimental picture will hopefully be clarified soon. 
Forming \mbox{nevertheless} the weighted averages of (\ref{Adir-exp}) and 
(\ref{Amix-exp}), using the rules of the Particle Data Group (PDG), yields
\begin{eqnarray}
{\cal A}_{\rm CP}^{\rm dir}(B_d\to\pi^+\pi^-)&=&-0.57\pm0.19 \,\, (0.32)
\label{Bpipi-CP-averages}\\
{\cal A}_{\rm CP}^{\rm mix}(B_d\to\pi^+\pi^-)&=&0.57\pm0.25 \,\, (0.61),
\label{Bpipi-CP-averages2}
\end{eqnarray}
where the errors in brackets are the ones increased by the PDG 
scaling-factor procedure \cite{PDG}. Direct CP violation at this 
level would require large penguin contributions with large CP-conserving 
strong phases. A significant impact of penguins on $B_d\to\pi^+\pi^-$ is 
also indicated by data on $B\to\pi K,\pi\pi$ \cite{FlMa2}, as well as by 
theoretical considerations \cite{BBNS3,PQCD-appl}. Consequently, it is 
already evident that the penguin contributions to $B_d\to\pi^+\pi^-$ 
{\it cannot} be neglected.

Many approaches to deal with the penguin problem in the extraction 
of weak phases from the CP-violating $B_d\to\pi^+\pi^-$ observables 
were developed; for a selection, see 
\cite{BBNS3,FlMa2,RF-dubna,Bpipi-strategies}. In 
Subsection~\ref{ssec:BsKK}, we shall return to this issue by having a 
closer look at an approach using $B_s\to K^+K^-$ \cite{RF-BsKK}.

\section{``El Dorado'' for Hadron Colliders: $B_s$-Meson System}\label{sec:Bs}
\subsection{General Features}\label{subsec:Bs-features}
At the $e^+e^-$ $B$ factories operating at the $\Upsilon(4S)$ resonance, 
no $B_s$ mesons are accessible, since $\Upsilon(4S)$ states decay only 
to $B_{u,d}$-mesons, but not to $B_s$. On the other hand, the physics 
potential of the $B_s$ system is very promising for hadron machines, 
where plenty of $B_s$ mesons are produced. Consequently,
$B_s$ physics is in some sense the ``El Dorado'' for $B$ experiments at 
hadron colliders. There are important differences between the $B_d$ and 
$B_s$ systems:
\begin{itemize}
\item Within the SM, the $B^0_s$--$\overline{B^0_s}$ mixing 
phase probes the tiny angle $\delta\gamma$ in the unitarity triangle
shown in Fig.\ \ref{fig:UT} (b), and is hence negligibly small:
\begin{equation}
\phi_s=-2\delta\gamma=-2\lambda^2\eta={\cal O}(-2^\circ),
\end{equation}
whereas $\phi_d=2\beta={\cal O}(50^\circ)$.
\item A large $x_s\equiv\Delta M_s/\Gamma_s={\cal O}(20)$ is expected in 
the SM, whereas $x_d=0.775\pm0.012$. The present lower bound on 
$\Delta M_s$ is given as follows \cite{LEPBOSC}:
\begin{equation}\label{Ms-bound}
\Delta M_s>14.4\,\mbox{ps}^{-1} \,\, (95\% \,\,{\rm C.L.}).
\end{equation}
\item There may be a sizeable width difference 
$\Delta\Gamma_s/\Gamma_s={\cal O}(-10\%)$ between the mass eigenstates 
of the $B_s$ system, whereas $\Delta\Gamma_d$ is negligibly small 
\cite{BeLe}. The present CDF and LEP results imply \cite{LEPBOSC}
\begin{equation}
|\Delta\Gamma_s|/\Gamma_s<0.31 \,\, (95\% \,\,{\rm C.L.}).
\end{equation}
Interesting applications of $\Delta\Gamma_s$ are extractions of weak 
phases from ``untagged'' $B_s$ data samples, where we do not distinguish 
between initially present $B^0_s$ or $\overline{B^0_s}$ mesons, 
as argued in \cite{Bs-untagged}.
\end{itemize}

Let us discuss the r\^ole of $\Delta M_s$ for the determination
of the unitarity triangle in slightly more detail. The comparison of 
$\Delta M_d$ with $\Delta M_s$ allows an interesting determination 
of the side $R_t$ of the unitarity triangle. To this end, only a single 
$SU(3)$-breaking parameter $\xi$  is required, which measures 
$SU(3)$-breaking effects in non-perturbative mixing and decay parameters. 
It can be determined through lattice or QCD sum-rule calculations. 
The mass difference $\Delta M_s$ has not yet been measured. However, the
lower bounds on $\Delta M_s$ can be converted into upper bounds on $R_t$ 
through \cite{Buras-ICHEP96}
\begin{equation}\label{DMs-constr}
\left(R_t\right)_{\rm max}=0.83\times\xi\times\sqrt{\frac{15.0\,
{\rm ps}^{-1}}{\left(\Delta M_s\right)_{\rm min}}},
\end{equation}
already excluding a large part in the $\overline{\rho}$--$\overline{\eta}$ 
plane, and implying in particular $\gamma<90^\circ$. In a recent paper 
\cite{KroRy}, it is argued that $\xi$ may be significantly larger 
than the conventional range, $\xi=1.15\pm0.06 \to 1.32\pm0.10$. The excluded 
range in the $\overline{\rho}$--$\overline{\eta}$ plane would then be reduced,
shifting the upper limit for $\gamma$ closer to $90^\circ$. Hopefully, the 
status of $\xi$ will be clarified soon. In the near future, run II of the 
Tevatron should provide a measurement of $\Delta M_s$, thereby constraining 
the unitarity triangle and $\gamma$ in a much more stringent way.

\subsection{Benchmark Decay Modes of $B_s$ Mesons}
An interesting class of $B_s$ decays is due to $b(\overline{b})\to 
c \overline{u}s(\overline{s})$
quark-level transitions. Here we have to deal with pure tree decays, 
where both $B_s^0$ and $\overline{B_s^0}$ mesons may decay into the same 
final state $f$. The resulting interference effects between decay and 
mixing processes allow a {\it theoretically clean} extraction of 
$\phi_s+\gamma$ from 
\begin{equation}\label{xi-xi}
\xi_f^{(s)}\times\xi_{\overline{f}}^{(s)}=e^{-2i(\phi_s+\gamma)}.
\end{equation}
There are several well-known strategies on the market employing these 
features: we may consider the colour-allowed decays $B_s\to D_s^\pm K^\mp$ 
\cite{ADK}, or the colour-suppressed modes $B_s\to D^0\phi$ \cite{GL0}. In the
case of $B_s\to D_s^{\ast\pm} K^{\ast\mp}$ or $B_s\to D^{\ast0}\phi$,
the observables of the corresponding angular distributions provide
sufficient information to extract $\phi_s+\gamma$ from ``untagged''
analyses \cite{FD2}, requiring a sizeable $\Delta\Gamma_s$. A
``tagged'' strategy involving $B_s\to D_s^{\ast\pm} K^{\ast\mp}$ modes
was proposed in \cite{LSS-00}. Recently, strategies making use of
``CP-tagged'' $B_s$ decays were proposed \cite{FP}, which require a 
symmetric $e^+e^-$ collider operated at the $\Upsilon(5S)$ resonance. 
In this approach, initially present CP eigenstates $B_s^{\rm CP}$ are 
employed, which can be tagged through the fact that the 
$B_s^0/\overline{B_s^0}$ mixtures have anticorrelated CP eigenvalues 
at $\Upsilon(5S)$. Here we may use the transitions $B_s\to D_s^\pm K^\mp, 
D_s^\pm K^{\ast\mp}, D_s^{\ast\pm} K^{\ast\mp}$. Let us note that there 
is also an interesting counterpart of (\ref{xi-xi}) in the $B_d$ 
system \cite{BdDpi}, which employs $B_d\to D^{(\ast)\pm}\pi^\mp$ decays, 
and allows a determination of $\phi_d+\gamma$. 

The extraction of $\gamma$ from the phase $\phi_s+\gamma$ provided
by the $B_s$ approaches sketched in the previous paragraph requires 
$\phi_s$ as an additional input, which is negligibly small in the Standard 
Model. Whereas it appears to be quite unlikely that the pure tree decays 
listed above are affected significantly by new physics, as they involve 
no flavour-changing neutral-current processes, this is not the case for 
the $B^0_s$--$\overline{B^0_s}$ mixing phase $\phi_s$. In order to probe 
this quantity, the decay $B_s\to J/\psi\,\phi$, which is the counterpart
of $B_d\to J/\psi K_{\rm S}$, offers interesting strategies \cite{DFN,DDF1}. 
In contrast to $B_d\to J/\psi K_{\rm S}$, the final state of 
$B_s\to J/\psi\phi$ is an admixture of different CP eigenstates. In order 
to disentangle them, we have to use the angular distribution of the 
$J/\psi\to \ell^+\ell^-$ and $\phi\to K^+K^-$ decay products \cite{DDLR}. 
The corresponding observables are governed by \cite{LHC-BOOK}
\begin{equation}\label{Bspsiphi-obs}
\xi^{(s)}_{\psi\phi}\,\propto\, e^{-i\phi_s}
\left[1-2\,i\,\sin\gamma\times{\cal O}(10^{-3})\right].
\end{equation}
Since we have $\phi_s={\cal O}(-2^\circ)$ in the SM, the extraction of 
$\phi_s$ from the $B_s\to J/\psi[\to\ell^+\ell^-] \phi[\to K^+K^-]$ 
angular distribution may well be affected by hadronic uncertainties
at the $10\%$ level. These hadronic uncertainties, which may become an
important issue in the LHC era \cite{LHC-BOOK}, can be controlled through 
$B_d\to J/\psi\, \rho^0$, exhibiting some other interesting 
features \cite{RF-ang}. Since $B_s\to J/\psi\phi$ shows small CP-violating 
effects in the SM because of (\ref{Bspsiphi-obs}), this mode represents 
a sensitive probe to search for new-physics contributions to 
$B^0_s$--$\overline{B^0_s}$ mixing \cite{NiSi}. For a detailed discussion 
of ``smoking-gun'' signals of a sizeable value of $\phi_s$, see \cite{DFN}. 
There, also methods to determine this phase {\it unambiguously} are proposed.

\subsection{The $B_s\to K^+K^-$, $B_d\to\pi^+\pi^-$ System}\label{ssec:BsKK}
Since $B_s\to K^+K^-$ and $B_d\to\pi^+\pi^-$ are related to each other
through an interchange of all down and strange quarks, the $U$-spin 
flavour symmetry of strong interactions allows us to express the four 
observables ${\cal A}_{\rm CP}^{\rm dir}(B_s\to K^+K^-)$, 
${\cal A}_{\rm CP}^{\rm mix}(B_s\to K^+K^-)$, 
${\cal A}_{\rm CP}^{\rm dir}(B_d\to\pi^+\pi^-)$, 
${\cal A}_{\rm CP}^{\rm mix}(B_d\to\pi^+\pi^-)$ as functions of
two hadronic penguin parameters $d$ and $\theta$, as well as $\gamma$,
$\phi_d$ and $\phi_s$, which is negligibly small in the SM. Consequently,
$d$, $\theta$, $\gamma$, $\phi_d$ can be determined. If $\phi_d$
is fixed through $B_d\to J/\psi K_{\rm S}$, the use of the $U$ spin symmetry
in the extraction of $\gamma$ and the hadronic parameters can be minimized
\cite{RF-BsKK}. The approach has certain theoretical advantages, and
is also very promising from an experimental point of view. At run II of
the Tevatron and the LHC, one expects experimental accuracies for $\gamma$ of 
${\cal O}(10^\circ)$ \cite{TEV-BOOK} and ${\cal O}(1^\circ)$ \cite{LHC-BOOK}, 
respectively. For a collection of other $U$-spin strategies, see 
\cite{RF-BdsPsiK,RF-ang,U-spin-strat}.

\begin{figure}[t]
$$\hspace*{-1.cm}
\epsfysize=0.2\textheight
\epsfxsize=0.3\textheight
\epsffile{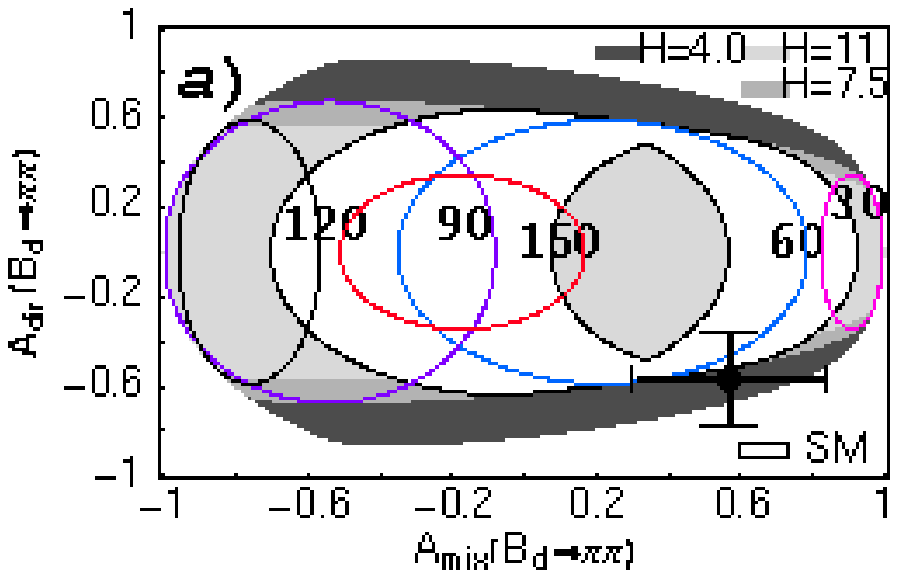} \hspace*{0.3cm}
\epsfysize=0.2\textheight
\epsfxsize=0.3\textheight
\epsffile{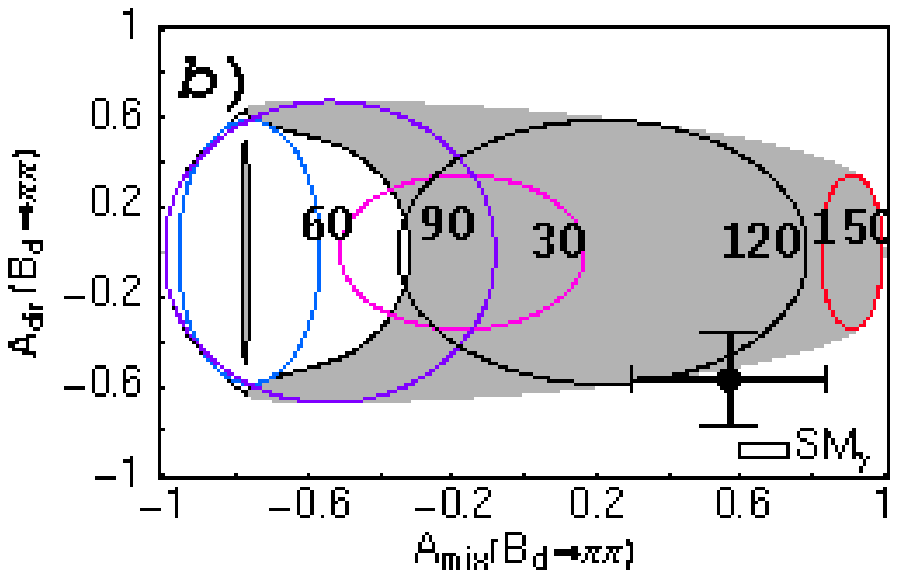}
$$
\caption[]{Allowed regions in $B_d\to\pi^+\pi^-$ observable space for 
(a) $\phi_d=47^\circ$ and various values of $H$, and (b) 
$\phi_d=133^\circ$ ($H=7.5$). The SM regions appear if we restrict 
$\gamma$ to (\ref{UT-Fit-ranges}). Contours representing fixed values 
of $\gamma$ are also included.}\label{fig:AdAmpipi}
\end{figure}

Since $B_s\to K^+K^-$ is not accessible at the $e^+e^-$ $B$ factories 
operating at $\Upsilon(4S)$, data are not yet available. However,  
$B_s\to K^+K^-$ is related to $B_d\to\pi^\mp K^\pm$ through an 
interchange of spectator quarks. Consequently, we may approximately
replace $B_s\to K^+K^-$ by $B_d\to\pi^\mp K^\pm$ to deal with the
penguin problem in $B_d\to\pi^+\pi^-$ \cite{U-variant}. To this end, 
the quantity
\begin{equation}\label{H-det}
H=\frac{1}{\epsilon}\left(\frac{f_K}{f_\pi}\right)^2
\left[\frac{\mbox{BR}(B_d\to\pi^+\pi^-)}{\mbox{BR}(B_d\to\pi^\mp K^\pm)}
\right]=
\left\{\begin{array}{ll}
7.3\pm2.9 & \mbox{(CLEO \cite{CLEO-BpiK})}\\
7.6\pm1.2 & \mbox{(BaBar \cite{BaBar-BpiK})}\\
7.1\pm1.9 & \mbox{(Belle \cite{Belle-BpiK})}
\end{array}\right.
\end{equation}
is particularly useful, where $\epsilon\equiv\lambda^2/(1-\lambda^2)$. 
It allows us to eliminate the hadronic parameter $d$ in
${\cal A}_{\rm CP}^{\rm dir}(B_d\to\pi^+\pi^-)$ and 
${\cal A}_{\rm CP}^{\rm mix}(B_d\to\pi^+\pi^-)$, which then depend -- 
for a given value of $\phi_d$ -- only on $\gamma$ and the strong phase 
$\theta$. If we vary $\gamma$ and $\theta$ within their allowed ranges, we 
obtain an allowed region in the ${\cal A}_{\rm CP}^{\rm dir}(B_d\to
\pi^+\pi^-)$--${\cal A}_{\rm CP}^{\rm mix}(B_d\to\pi^+\pi^-)$ plane
\cite{FlMa2}, which is shown in Fig.~\ref{fig:AdAmpipi}. We observe that 
the experimental averages (\ref{Bpipi-CP-averages}) and 
(\ref{Bpipi-CP-averages2}), represented by the crosses, overlap nicely with 
the SM region for $\phi_d=47^\circ$, and point towards $\gamma\sim55^\circ$. 
In this case, not only $\gamma$ would be in accordance with the results 
of the CKM fits (\ref{UT-Fit-ranges}), but also $\phi_d$. On the other hand, 
for $\phi_d=133^\circ$, the experimental values favour $\gamma\sim125^\circ$, 
and have essentially no overlap with the SM region. Since a value of 
$\phi_d=133^\circ$ would require CP-violating new-physics contributions 
to $B^0_d$--$\overline{B^0_d}$ mixing, also the $\gamma$ range in 
(\ref{UT-Fit-ranges}) may no longer hold in this case, as it relies on a 
Standard-Model interpretation of the experimental information on 
$B^0_{d,s}$--$\overline{B^0_{d,s}}$ mixing. In particular, also values 
for $\gamma$ larger than $90^\circ$ could then in principle be accommodated. 
As discussed in detail in \cite{FlMa2}, in order to put these statements
on a more quantitative basis, we may use $H$ to calculate 
${\cal A}_{\rm CP}^{\rm dir}(B_d\to\pi^+\pi^-)$ for given values of
${\cal A}_{\rm CP}^{\rm mix}(B_d\to\pi^+\pi^-)$ as a function of $\gamma$.
Taking into account (\ref{Bpipi-CP-averages}) and (\ref{Bpipi-CP-averages2}),
we then obtain
\begin{equation}\label{gam-res}
34^\circ\lsim\gamma\lsim75^\circ \, (\phi_d=47^\circ), \quad
105^\circ\lsim\gamma\lsim146^\circ \, (\phi_d=133^\circ).
\end{equation}
In the future, the experimental uncertainties of the $B_d\to\pi^+\pi^-$
observables will be reduced considerably, thereby providing significantly 
more stringent results for $\gamma$, as well as the hadronic parameters. 

\begin{figure}[t]
$$\hspace*{-1.cm}
\epsfysize=0.2\textheight
\epsfxsize=0.3\textheight
\epsffile{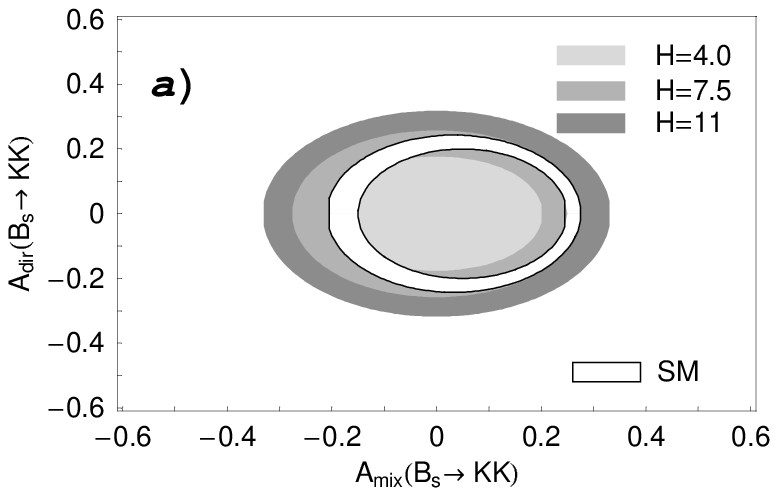} \hspace*{0.3cm}
\epsfysize=0.2\textheight
\epsfxsize=0.3\textheight
 \epsffile{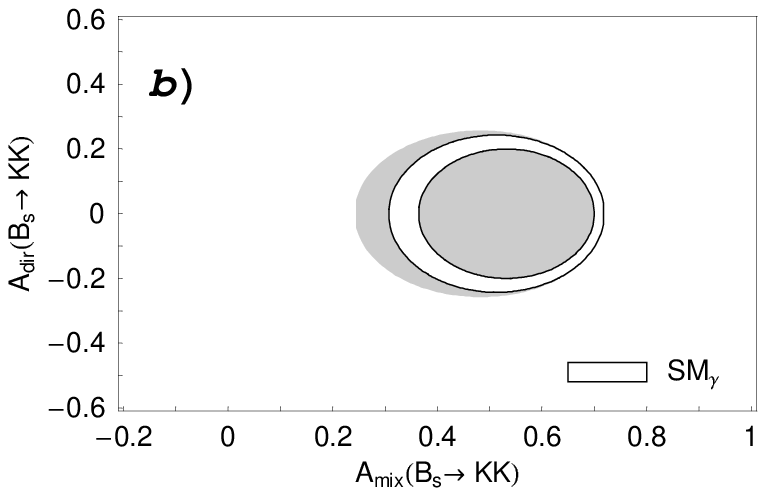}
$$
\caption[]{Allowed regions in $B_s\to K^+K^-$ observable space for (a) 
$\phi_s=0^\circ$ and various values of $H$, and (b) 
$\phi_s^{\rm NP}=30^\circ$ ($H=7.5$). The SM regions appear if 
$\gamma$ is restricted to (\ref{UT-Fit-ranges}).}\label{fig:Ams-Ads}
\end{figure}

In analogy to the $B_d\to\pi^+\pi^-$ analysis discussed above, we may also 
use $H$ to obtain an allowed region in the ${\cal A}_{\rm CP}^{\rm mix}(B_s\to 
K^+K^-)$--${\cal A}_{\rm CP}^{\rm dir}(B_s\to K^+K^-)$ plane \cite{FlMa2},
as shown in Fig.~\ref{fig:Ams-Ads}. There, also the impact of a non-vanishing 
value of $\phi_s$, which may be due to new-physics contributions to 
$B^0_s$--$\overline{B^0_s}$ mixing, is illustrated. If we constrain $\gamma$ 
to (\ref{UT-Fit-ranges}), even more restricted regions appear. The allowed 
regions are remarkably stable with respect to variations of parameters 
characterizing $U$-spin-breaking effects \cite{FlMa2}, and represent a 
narrow target range for run II of the Tevatron and the experiments of the 
LHC era, in particular LHCb and BTeV. These experiments will allow us to 
exploit the whole physics potential of the $B_d\to\pi^+\pi^-$, 
$B_s\to K^+K^-$ system \cite{RF-BsKK}.

\section{Comments on Rare $B$ Decays}\label{sec:rare}
Let us finally comment briefly on rare $B$ decays, which occur at the 
one-loop level in the SM, and involve $\overline{b}\to\overline{s}$ or 
$\overline{b}\to\overline{d}$ flavour-changing neutral-current transitions. 
Prominent examples are $B\to K^\ast\gamma$, $B\to \rho\gamma$, 
$B\to K^\ast\mu^+\mu^-$ and $B_{s,d}\to \mu^+\mu^-$. Within the SM, 
these modes exhibit small branching ratios at the $10^{-4}$--$10^{-10}$ 
level, and do not -- apart from $B\to \rho\gamma$ -- show sizeable 
CP-violating effects, thereby representing important probes to search 
for new physics. For detailed discussions of the many interesting aspects 
of rare $B$ decays, the reader is referred to the overview articles 
listed in \cite{rare}.

\section{Conclusions and Outlook}\label{sec:concl}
Decays of $B$ mesons represent a very exciting field of research. Thanks 
to the efforts of the BaBar and Belle collaborations, CP violation could 
recently be established in the $B$ system with the help of the 
``gold-plated'' mode $B_d\to J/\psi K_{\rm S}$, thereby opening a new 
era in the exploration of CP violation. The world average 
$\sin2\beta=0.734\pm0.054$ now agrees well with the SM expectation, but 
leaves a twofold solution $\phi_d\sim 47^\circ \lor 133 ^\circ$ for the 
$B^0_d$--$\overline{B^0_d}$ mixing phase itself. As the latter solution
would point towards new physics, it is important to resolve this ambiguity 
directly.

The physics potential of $B$ experiments goes far beyond 
$B_d\to J/\psi K_{\rm S}$, allowing us now to confront many more CP
strategies with data. Here the main goal is to overconstrain the unitarity 
triangle as much as possible, where $B\to\pi K$, $B\to\phi K$ and
$B\to\pi\pi$ are important benchmark modes. Studies of $B$ decays at 
hadron colliders are an essential element of this programme, providing --
among other things -- access to the $B_s$-meson system. Already run II of 
the Tevatron is expected to yield interesting results on $B_s$ physics, 
and should discover $B^0_s$--$\overline{B^0_s}$ mixing soon, which is an 
important ingredient for the ``standard'' analysis of the unitarity triangle. 
Prominent $B_s$ decays are $B_s\to J/\psi \phi$, $B_s\to K^+K^-$ and 
$B_s\to D^\pm_s K^\mp$. Although we may obtain first valuable insights
into these modes at the Tevatron, they can only be fully explored at 
the experiments of the LHC era, in particular LHCb and BTeV.

%INDEX%%%%%%%%%%%%%%%%%%%%%%%%%%%%%%%%%%%%%%%%%%%%%%%%%%%%%%%%%%%%%%%
% Please check with the editor of your book whether he plans to
% include a "mutual" subject index - if so, please code your entries
% in the standard syntax. For your own purposes you may print your
% "personal" index by using the following commands:
%
%\clearpage
%\addcontentsline{toc}{section}{Index}
%\flushbottom
%\printindex
%%%%%%%%%%%%%%%%%%%%%%%%%%%%%%%%%%%%%%%%%%%%%%%%%%%%%%%%%%%%%%%%%%%%%

\end{document}